\documentclass[runningheads]{llncs}
\usepackage{graphicx}
\usepackage{changes}
\usepackage{hyperref}
\usepackage{url}
\usepackage{booktabs}
\usepackage{subcaption}
\usepackage{tikz}

\usepackage{bm}
\usepackage{color}
\usepackage{amssymb}

\usepackage{algorithmic}
\usepackage{algorithm}
\usepackage[cmex10]{amsmath}
\usepackage{url}
\usepackage{float}

\usepackage{multirow}

\def\R{{\mathbb R}}

\def\p{{\mathrm{p}}}

\def\bff{{\mathbf f}}

\def\bx{{\mathbf x}}

\def\bz{{\mathbf z}}

\def\bA{{\mathbf A}}

\def\bD{{\mathbf D}}

\def\bL{{\mathbf L}}

\def\bX{{\mathbf X}}

\def\bZ{{\mathbf Z}}

\newcommand{\CNN}{\mathrm{CNN}}
\newcommand{\Att}{\mathrm{Att}}
\newcommand{\CE}{\mathrm{CE}}

\definechangesauthor[name={Yunan}, color=red]{as}

\begin{document}

\title{Smooth Attention for Deep Multiple Instance Learning: Application to CT Intracranial Hemorrhage Detection}
\titlerunning{SA-DMIL}
%
\author{Yunan Wu\inst{1} \and
Francisco M. Castro-Macías\inst{2,4} \and
Pablo Morales-Álvarez\inst{3,4} \and
Rafael Molina \inst{2} \and
Aggelos K. Katsaggelos \inst{1}}

%
\authorrunning{Y.Wu et al.}
%

\institute{Image and Video Processing Laboratory, Department of Electrical and Computer 
Engineering, Northwestern University, USA
\and
Department of Computer Science and Artificial Intelligence, University of Granada, Spain
\and
Department of Statistics and Operations Research, University of Granada, Spain
\and 
Research Centre for Information and Communication Technologies (CITIC-UGR), University of Granada, Spain
}

%


\maketitle  

\vspace{-2mm} 
\begin{abstract}
Multiple Instance Learning (MIL) has been widely applied to medical imaging diagnosis, where bag labels are known and instance labels inside bags are unknown. Traditional MIL assumes that instances in each bag are independent samples from a given distribution. However, instances are often spatially or sequentially ordered, and one would expect similar diagnostic importance for neighboring instances. To address this, in this study, we propose a smooth attention deep MIL (SA-DMIL) model. Smoothness is achieved by the introduction of first and second order constraints on the latent function encoding the attention paid to each instance in a bag. The method is applied to the detection of intracranial hemorrhage (ICH) on head CT scans.
The results show that this novel SA-DMIL: (a) achieves better performance than the non-smooth attention MIL at both scan (bag) and slice (instance) levels; (b) learns spatial dependencies between slices; and (c) outperforms current state-of-the-art MIL methods on the same ICH test set.   
\end{abstract}




\vspace{-8mm}
\section{Introduction}
\vspace{-1mm} 

Multiple Instance Learning (MIL) \cite{quellec2017multiple,CARBONNEAU2018329} is a type of weakly supervised learning that has become very popular in biomedical imaging diagnostics due to the reduced annotation effort it requires \cite{cheplygina2019not,gadermayr2022multiple}. 
In the case of MIL binary classification, the training set is partitioned into bags of instances. 
Both bags and instances have labels, but only bag labels are observed while instance labels remain unknown. 
It is assumed that a bag label is positive if and only if the bag contains at least one positive instance \cite{dietterich1997solving}. 
The goal is to produce a method that, trained on bag labels only, is capable of predicting both bag and instance labels.

Among the proposed approaches for learning in the MIL scenario \cite{quellec2017multiple}, deep learning (DL) methods stand out when dealing with highly structured data (such as medical images and videos) \cite{lecun2015deep}.
The most successful deep MIL approaches combine an instance-level processing mechanism (i.e., a feature extractor) with a pooling mechanism to aggregate information from instances in a bag \cite{cheplygina2019not,gadermayr2022multiple}. 
Among the pooling operators, the attention-based weight pooling proposed in \cite{ilse2018attention} is frequently used as a way to discover \emph{key instances}, i.e., those responsible for the label of a bag. 
However, this pooling operator was formulated under strong assumptions of independence between the instances in a bag. 
This is a drawback in biomedical imaging problems, where instances in a bag are often spatially or sequentially ordered and their diagnostic importance is expected to be similar for neighboring instances \cite{shao2021transmil,li2021multi}.

In this work, we are particularly interested in the detection of intracranial hemorrhage (ICH), a serious life-threatening emergency caused by blood leakage inside the brain \cite{caceres2012intracranial,qureshi2001spontaneous}. 
Radiologists confirm the presence of ICH by using computed tomography (CT) scans \cite{cordonnier2018intracerebral}, which consist of a significant number of slices, each representing a section of the head at a given height.
Unfortunately, the shortage of specialized radiologists and their increasing workload sometimes lead to delayed and erroneous diagnoses \cite{arendts2003cranial,erly2002radiology,strub2007overnight,mcdonald2015effects}, which may result in potentially preventable cerebral injury or morbidity \cite{elliott2010acute,cordonnier2018intracerebral}.
For this reason, there is a growing interest in the development of automated systems to assist radiologists in making rapid and reliable diagnoses.

State-of-the-art ICH detection methods rely on DL models, specifically convolutional neural networks (CNNs), to extract meaningful ICH features \cite{yeo2021review}. 
However, 2D CNNs need to be coupled with other mechanisms such as recurrent neural networks (RNNs) \cite{ye2019precise,grewal2018radnet} or 3D CNNs \cite{chang2018hybrid,ker2019image,titano2018automated,arbabshirani2018advanced} to account for interslice dependencies. 
Although these approaches are quite successful in terms of performance, their use is limited by the large amount of labeled data they require \cite{yeo2021review}.
To address this issue, the ICH detection task has been formulated as an MIL problem, achieving comparable performance to fully supervised models while reducing the workload of radiologists \cite{wu2021combining,teneggi2022weakly}. Note that the MIL framework is naturally suited for the ICH detection problem since a CT scan (i.e., a bag) is considered positive if it contains at least one slice (i.e., an instance) with evidence of hemorrhage (i.e., positive instance).


In this work, we improve upon the state-of-the-art deep MIL methods by introducing dependencies between instances in a sound probabilistic manner. These dependencies are formulated over a neighborhood graph to impose smoothness on the latent function that encodes the attention given to each instance. Smoothness is achieved by introducing specific first- and second-order constraints on the latent function. Our model, called SA-DMIL, is applied to the ICH detection problem, obtaining (a) significant improvements upon the performance of non-smooth models at both scan and slice levels, (b) smoother attention weights across slices by benefiting from the inter-slice dependencies, and (c) a superior performance against other popular MIL methods on the same test set. 
\section{Methods}
\vspace{-1mm} 

\vspace{-1mm}
\subsection{Problem formulation}
\vspace{-1mm} 

We start by formulating ICH detection as a Multiple Instance Learning (MIL) problem. To do so, we map slices to instances and CT scans to bags. The slices (instances) will be denoted by $\bx_i^b \in \R^{3HW}$, where $H$ and $W$ are the height and width of the image, $3$ is the number of color channels, $b$ is the index of the scan to which the slice belongs to and $i$ is the index of the slice inside the bag. We will denote the label of a slice by $y_i^b \in \left\{ 0,1\right\}$. If the slice contains hemorrhage, then $y_i^b=1$, otherwise $y_i^b=0$. Note that the slice labels remain unknown since only scan labels are given. As we know, slices are grouped to form the CT scans. Each scan (bag) will be denoted by $\bX^b = \left[ \bx_1^b, \ldots, \bx_{N_b}^b \right]^\top \in \R^{N_b \times 3HW}$. Here, $N_b$ is the number of slices in bag $b$. We will assume that $B$ CT scans are given, so $b \in \left\{ 1, \ldots, B \right\}$. Given a CT scan $b$, we will denote its label by $T^b \in \left\{ 0,1\right\}$. Notice that $T^b = 1$ if and only if some of ${y_i}^b = 1$, i.e., the following relationship between scan and slice labels holds, 
\begin{equation}\label{eq:mil_assumption}
    T^b = \max \left\{ y_1^b, \ldots, y_{N_b}^b\right\}.
\end{equation}

\vspace{-5mm} 
\subsection{Attention-based Multiple Instance Learning pooling}
\vspace{-1mm} 

The attention-based MIL pooling was proposed in \cite{ilse2018attention} as a way to discover \emph{key instances}, i.e., those responsible for the diagnosis of a scan. It consists of a weighted average of instances (low-dimensional embeddings) where the weights are parameterized by a neural network. Formally, given a bag of $N_b$ embeddings $\bZ^b = \left[ \bz_1^b, \ldots, \bz_{N_b}^b \right]^\top$, where $\bz_i^b \in \R^{D}$, the attention-based MIL pooling computes 
\begin{equation}
    \Phi_{\Att}\left( \bZ^b \right) = \textstyle\sum_{i=1}^{N_b} s(\bz_i^b) \bz^b_i,
\end{equation}
where 
\begin{equation}\label{eq:att_def}
    s\left(\bz_i^b\right)=\frac{\exp \left(f\left(\bz_i^b\right)\right)}{\sum_{j}^{N_b} \exp \left(f\left(\bz_j^b\right)\right)}, \quad f\left(\bz_i^b\right) = \mathbf{w}^{\top} \tanh \left(\mathbf{V} \mathbf{z}_{i}^b\right).
\end{equation}
Notice that $\mathbf{w} \in \mathbb{R}^{L}$ and $\mathbf{V} \in \mathbb{R}^{L \times D}$ are trainable parameters, where $D$ denotes the size of feature vectors. We refer to $s\left(\bz_i^b\right)$ as \emph{attention weights} and to $f\left(\bz_i^b\right)$ as \emph{attention values}. 

This operator  was proposed under the assumption that the instances in a bag show neither dependency nor order among each other. 
Although this may be the case in simple problems, it does not occur in problems such as ICH detection. 
Note that the attention weights of slices in a bag are correlated: given a slice containing ICH, we expect that the adjacent slices will also contain ICH with high probabilities. This is essential in finding slices with ICH. In the next subsection, we show how to introduce this correlation between attention weights.

\begin{figure}
\vspace*{-4mm}
\centering
\includegraphics[width=1.00\textwidth]{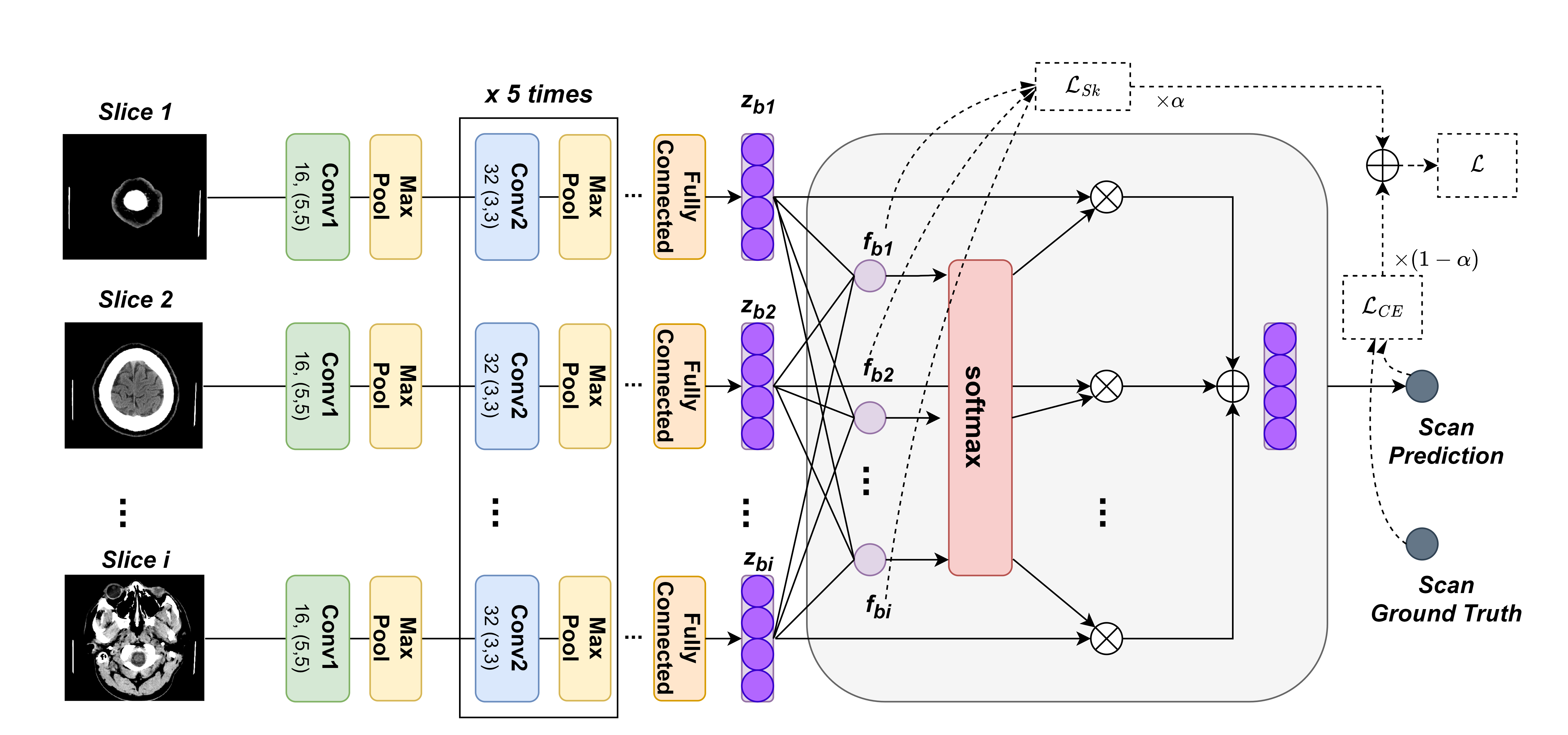}
\caption{SA-DMIL architecture. It consists of CNNs that extract slice level features and an attention block to aggregate slice features. The loss function is a weighted average of the binary cross entropy and a novel smooth attention loss.} \label{fig_model}
\vspace*{-5mm}
\end{figure}

\vspace*{-3mm} 
\subsection{Modeling correlation through the attention mechanism}\label{sec:methods_modelling_correlation}
\vspace*{-1mm} 

Ideally, in the case of a positive scan ($T^b = 1$), high attention weights should be assigned to slices that are likely to have a positive label ($y_i^b=1$). Given the dependency between slices, contiguous slices should have similar attention values. In other words, the differences between the attention values of contiguous slices should be \emph{small}. Thus, for each bag $b$, these quantities should be small 
\begin{gather}
    \mathcal{L}_{S1}^b = 2^{-1}\textstyle\sum_{\substack{i, j \in \operatorname{Bag}(b)}} A^b_{ij} \left( f\left(\bz_i^b\right) - f\left(\bz_{j}^b\right) \right)^2, \label{eq:L_S1_b}\\
    \mathcal{L}_{S2}^b = 4^{-1}\textstyle\sum_{i \in \operatorname{Bag}(b)} \left(\textstyle\sum_{\substack{j \in \operatorname{Bag}(b)}} A^b_{ij} \left(f\left(\bz_i^b\right) - f\left(\bz_j^b\right)\right)\right)^2, \label{eq:L_S2_b}
\end{gather}
where $A^b_{ij} = 1$ if the slices $i, j$ are related in bag $b$, and $0$ otherwise. We smooth $f\left(\bz_i^b\right)$ instead of $s\left(\bz_i^b\right)$ because a non-constrained parameter $f$ ensures consistent smoothing while $s$ requires a normalization across instances in a bag.

Equations \eqref{eq:L_S1_b} and \eqref{eq:L_S2_b} correspond, respectively, to the energies of the, so called, conditional and simultaneous autoregressive models in the statistics literature \cite{ripley1981spatial,belkin2006manifold}. 
For our problem, they model the value of $f$ at a given location (instance) given the values at neighboring instances. From the regularization viewpoint, these terms constrain the first and second derivatives of the function $f$, respectively, which favors smoother functions (examine the zero of the derivative of $f$). 
That is, a \textit{priori} all attention weights are expected to be the same because $f$ is expected to be constant. As observations arrive, they change to reflect the importance of each instance. 
Note that \eqref{eq:L_S1_b} and \eqref{eq:L_S2_b} impose smoothness but they can be modified to model, for example, competition between the attention weights by simply replacing the minus sign with a plus sign.

To compute $\mathcal{L}_{S1}^b$ and $\mathcal{L}_{S2}^b$ efficiently we consider the simple graph defined by the dependency between slices. For a bag $b$, its adjacency matrix is $\bA^b = \left[ A_{ij}^b \right]$. The degree matrix $\bD^b = \left[ D_{ij}^b \right]$ is a diagonal matrix that contains the degree of each slice (the degree of the slice $i$ is the number of slices $j$ such that $A_{ij}^b=1$). This is, $D_{ii}^b = \mathrm{degree}(i)$ and $D_{ij}^b = 0$ if $i \neq j$. Using these, one can compute the graph Laplacian matrix of a bag as $\bL^b = \bD^b - \bA^b$. It is easy to show that 
\begin{equation}
    \mathcal{L}_{S1}^b = {\bff^b}^\top \bL^b \bff^b, \quad \mathcal{L}_{S2}^b = {\bff^b}^\top \bL^b \bL^b \bff^b,
\end{equation}
where $\bff^b = \left[ f(\bz_1^b), \ldots, f(\bz_{N_b}^b)\right]^\top$. 
The sum of $\mathcal{L}_{Sk}^b$ over bags, where $k \in \left\{ 1,2\right\}$, can be added to the loss function of a network to be minimized along the task-specific loss. 
Note that these two terms provide two different approaches to exploiting the correlations between instances through the loss function. 
We will refer to this approach as smooth attention (SA) loss. 
In the following subsection, we propose a model that can use either $\mathcal{L}_{S1}$ or $\mathcal{L}_{S2}$. The effect of each term will be discussed in section \ref{sec:results}.

\vspace*{-3mm} 
\subsection{SA-DMIL model description}
\vspace*{-1mm} 

We propose to couple the attention-based MIL pooling with the SA loss terms introduced in subsection \ref{sec:methods_modelling_correlation}. The proposed model, named Smooth Attention Deep Multiple Instance Learning (SA-DMIL), is depicted in Fig. \ref{fig_model}. We use a Convolutional Neural Network (CNN), denoted by $\Phi_{\CNN}$, as a feature extractor to obtain a vector of low dimensional embeddings for each instance. That is, given a bag $\bX^b = \left[ \bx_1^b, \ldots, \bx_{N_b}^b \right]$, where $\bx_n^b \in \mathbb{R}^{3\times HW}$, we compute
\begin{gather}
    \bz_n^b = \Phi_{\CNN}\left( \bx_n^b \right) \in \R^{D}, \quad \bZ^b = \left[ \bz_1^b, \ldots, \bz_{N_b}^b \right].
\end{gather}
The CNN module in Fig. \ref{fig_model} is implemented with six convolutional blocks, followed by a flatten layer. 
$\bZ^b$ is then fed into the attention layer $\Phi_{\Att}$ described in subsection \ref{sec:methods_modelling_correlation} to obtain a scan representation. After that, the scan representation passes through a classifier $\Phi_\mathrm{c}$ (i.e., one fully connected layer with a sigmoid activation) to predict the scan labels,
\begin{equation}
    \p\left(T^b \mid \bX^b\right) \approx \Phi\left( \bX^b \right) =  \Phi_\mathrm{c}\left( \Phi_{\Att}\left( \Phi_{\CNN} \left(\bX^b\right) \right) \right),
\end{equation}
where we have written $\Phi_{\CNN} \left(\bX^b\right) = \left[ \Phi_{\CNN}\left(\bx_{1}^b\right), \ldots, \Phi_{\CNN}\left(\bx_{N_b}^b\right) \right]$. 
Our model, that corresponds to the composition $\Phi = \Phi_\mathrm{c} \circ \Phi_\Att \circ \Phi_\CNN$, is trained using the following loss function until convergence, 
\begin{equation}
    \mathcal{L} = \left(1-\alpha\right)\mathcal{L}_{\CE} + \alpha\mathcal{L}_{Sk},
\end{equation}
where $\alpha \in \left[0,1\right]$ is an hyperparameter and $\mathcal{L}_{\CE}$ the common cross-entropy loss,
\begin{equation}
    \mathcal{L}_{\CE} = \textstyle\sum_b \left[ T^b \log\left( \Phi\left(\bX^b\right) \right) + \left(1-T^b\right)\log\left( 1-\Phi\left(\bX^b\right) \right)\right],
\end{equation}
where $k\in \{1,2\}$, and $\mathcal{L}_{Sk} = \sum_b \mathcal{L}_{Sk}^b$ (see equations \eqref{eq:L_S1_b} and \eqref{eq:L_S2_b}). 
Depending on the value of $k$, we obtain two variations of SA-DMIL, which will be referred to as SA-DMIL-$S1$ and SA-DMIL-$S2$. 
The baseline model, Att-MIL (non-smooth attention), is recovered when $\alpha=0.0$ \cite{ilse2018attention}.
Following the approach of previous studies \cite{wu2021combining,lopez2022deep}, attention weights will be used to obtain predictions at the slice level (although they are not specifically designed for it). 
If a scan is predicted to be negative, all slices are also predicted to be negative, while if a scan is predicted to correspond to an ICH, slices whose attention weight is above a threshold (i.e., $1/N_b$, with $N_b$ being the number of slices in that scan) are predicted as ICH. 

\section{Experimental design}\label{sec:experimental_design}
\vspace*{-1mm} 
\subsection{Data and data preprocessing}
\vspace*{-1mm} 
The dataset used in this work was obtained from the 2019 Radiological Society of North America (RSNA) challenge \cite{noauthor_rsna_nodate}, which included 39650 CT slices from 1150 subjects. 
The data were split among subjects, with 1000 scans (ICH: Normal scans = 411: 589; ICH: Normal slices = 4976: 29520) used for training and validation, and the remaining 150 scans (ICH: Normal scans = 72: 78; ICH: Normal slices = 806: 4448) used for held-out testing. 
The number of slices in the scans varied from 24 to 57. 
All CT slices underwent the same preprocessing procedure as described in \cite{wu2021combining}. 
Each CT slice had three windows applied to its original Hounsfield Units by changing the window Width (W) and Center (C) to manipulate the display of specific tissues, as radiologists typically do when diagnosing brain CTs.
Here, we selected the brain (W: 80, C:40), subdural (W:200, C:80) and soft tissue (W:380, C: 40) windows. 
All images were then resized to the same size of $512 \times 512$ and normalized to the range $[0,1]$. 
CTs were annotated at both the scan and slice levels, but slice labels were used for evaluation only, while scan labels were used for training and evaluation.

\vspace*{-3mm}
\subsection{Experimental settings} \label{sec: experimental_setting}
\vspace*{-1mm} 
We fix $D=128$ and $L=50$ in equation \eqref{eq:att_def}. We use the Adam optimizer with the learning rate starting at $10^{-4}$. The batch size is set to 4, the maximum number of epochs is set to 200 and the patience for early stopping is set to 8. We test different values of the $\alpha$ hyperparameter, between 0 and 1 with a jump of 0.1. All experiments were run 5 independent times and the mean and standard deviation were reported in the held-old testing set at both scan and slice levels. The average training time is 10.3 hours for SA-DMIL-$S1$ and 10.5 hours for SA-DMIL-$S2$. The prediction time is approximately 15.8 seconds for each scan. All experiments were conducted using Tensorflow 2.11 in Python 3.8 on a single GPU (NVIDIA Quadro RTX 8000). The code will be available via \href{https://github.com/YunanWu2168/SA-MIL}{GitHub}.      
\vspace*{-1mm} 

\section{Results and discussion} \label{sec:results}
\vspace{-2mm} 
\subsection{Hyperparameters tuning}
\vspace{-1mm} 
In this subsection, we study the effect of SA loss in terms of performance. 
Table \ref{table:reduced} compares the performance of models for different values of $\alpha$. 
The standard deviation and other values of $\alpha$ can be found in the appendix, Tables S1 and S2. The results show that at both scan and slice levels, adding a smoothness term to the loss function ($\alpha > 0.0$) achieves better performance than Att-MIL ($\alpha=0.0$). 
These improvements are significant, with increases in accuracy, F1 and AUC scores of approximately 7\%, 9\% and 5\% respectively, at scan level, and increases in accuracy and F1 score of 8\% and 11\% respectively, at slice level. 
The recall is the only metric in which our model does not excel, where the baseline Att-MIL obtains the best value. However, this is associated with very low precision values. 
Note that, as $\alpha$ increases, the performance of the model first improves and then drops, which is consistent with the role played by the SA loss as a regularization term. The difference between $\mathcal{L}_{S1}$ and $\mathcal{L}_{S2}$ is not significant although $\mathcal{L}_{S1}$ performs slightly better.
In fact, when using $\mathcal{L}_{S1}$, $\alpha = 0.5$ gives the best diagnostic performance with an AUC of 0.879 ($\pm$ 0.003) at scan level and an accuracy of 0.834 ($\pm$ 0.010) at slice level. 

\begin{table}[ht]
\vspace*{-0.65cm}
\scriptsize
\caption{Performance of SA-DMIL and other MIL methods at slice and scan levels on the RSNA dataset. The average of 5 independent runs is reported. For space constraints, the standard deviation is reported in the appendix.}
\label{table:reduced}
\centering
\resizebox{\textwidth}{!}{%
\begin{tabular}{@{}cc|ccccc|cccc@{}}
\toprule
\multicolumn{2}{c|}{} & \multicolumn{5}{c|}{Scan level} & \multicolumn{4}{c}{Slice level} \\
\multicolumn{2}{c|}{Model} & Acc & Pre & Rec & F1 & AUC & Acc & Pre & Rec & F1 \\ \midrule 
\multirow{5}{*}{SA-DMIL-$S1$} & $\alpha=0.9$ & 0.753 & 0.803 & 0.681 & 0.735 & 0.839 & 0.789 & 0.670 & 0.541 & 0.598 \\
 & $\alpha=0.7$ & 0.806 & 0.763 & 0.784 & 0.775 & 0.860 & 0.828 & 0.679 & 0.576 & 0.639 \\
 & $\alpha=0.5$ & \textbf{0.813} & 0.805 & 0.806 & \textbf{0.806} & \textbf{0.879} & \textbf{0.834} & 0.732 & \textbf{0.608} & \textbf{0.686} \\
 & $\alpha=0.3$ & 0.767 & 0.734 & 0.806 & 0.768 & 0.859 & 0.775 & 0.702 & 0.551 & 0.624 \\
 & $\alpha=0.1$ & 0.747 & 0.783 & 0.652 & 0.712 & 0.841 & 0.766 & 0.649 & 0.540 & 0.584 \\ \midrule
\multirow{5}{*}{SA-DMIL-$S2$} & $\alpha=0.9$ & 0.753 & 0.817 & 0.613 & 0.714 & 0.816 & 0.768 & 0.733 & 0.551 & 0.598 \\
 & $\alpha=0.7$ & 0.767 & 0.776 & 0.722 & 0.748 & 0.843 & 0.807 & 0.734 & 0.591 & 0.638 \\
 & $\alpha=0.5$ & 0.800 & 0.828 & 0.736 & 0.780 & 0.867 & 0.823 & \textbf{0.748} & 0.596 & 0.659 \\
 & $\alpha=0.3$ & 0.763 & 0.797 & 0.686 & 0.721 & 0.853 & 0.790 & 0.738 & 0.561 & 0.622 \\
 & $\alpha=0.1$ & 0.747 & 0.736 & 0.740 & 0.736 & 0.833 & 0.767 & 0.683 & 0.547 & 0.593 \\ \midrule 
\multicolumn{2}{c|}{Att-MIL ($\alpha = 0.0$) \cite{ilse2018attention}} & 0.740 & 0.674 & 0.832 & 0.719 & 0.829 & 0.751 & 0.623 & 0.543 & 0.579 \\
\multicolumn{2}{c|}{MIL + Max agg. \cite{wang2019comparison}} & 0.617 & \textbf{0.856} & 0.447 & 0.575 & 0.743 & 0.732 & 0.441 & 0.373 & 0.406 \\
\multicolumn{2}{c|}{MIL + Mean agg. \cite{wang2019comparison}} & 0.677 & 0.670 & 0.734 & 0.693 & 0.801 & 0.741 & 0.502 & 0.386 & 0.447 \\
\multicolumn{2}{c|}{Att-CNN + VGPMIL \cite{wu2021combining}} & 0.765 & 0.724 & \textbf{0.851} & 0.773 & 0.868 & 0.807 & 0.714 & 0.538 & 0.597 \\ \bottomrule
\end{tabular}%
}
\vspace*{-5mm}
\end{table}

\vspace*{-3mm} 
\subsection{Smooth Attention MIL vs. other MIL methods}
\vspace*{-2mm} 
The performance of other popular MIL methods is also included in Table \ref{table:reduced}. 
All method share the same CNN architecture to extract slice features, but they differ in the pooling operator they use: Max \cite{wang2019comparison}, Mean \cite{wang2019comparison}, Attention \cite{ilse2018attention} or Gaussian Process (GP) \cite{wu2021combining}. 
These results show that the performance of SA-DMIL is consistently better than other methods across different metrics and at both scan and slice levels. 
Only the precision of MIL+Max agg. and the recall of AttCNN+VGPMIL at scan level are higher than those obtained by SA-DMIL. 
However, considering the trade-off between precision and recall given by F1, our method achieves a superior performance. 
In tasks like ICH detection, where neighbouring instances are expected to have similar diagnostic importance. Unlike other MIL methods that assume each instance to be independently distributed, SA-DMIL stands out by considering the spatial correlation between instances, which compels it to learn more meaningful features for making accurate bag predictions.
Notably, this is achieved by simply adding a smoothing term to the loss function without increasing the number of model parameters.
This can potentially be applied to existing architectures to further improve performance without adding complexity.    
    
\vspace*{-3mm} 
\subsection{Visualizing smooth regularizing effects at slice level}
\begin{figure}[ht]
\vspace*{-4mm} 
\centering
	\subfloat[Scan 1.]{
                \includegraphics[trim={0.4cm 1cm 2cm 1.3cm},clip,width=0.3\textwidth]{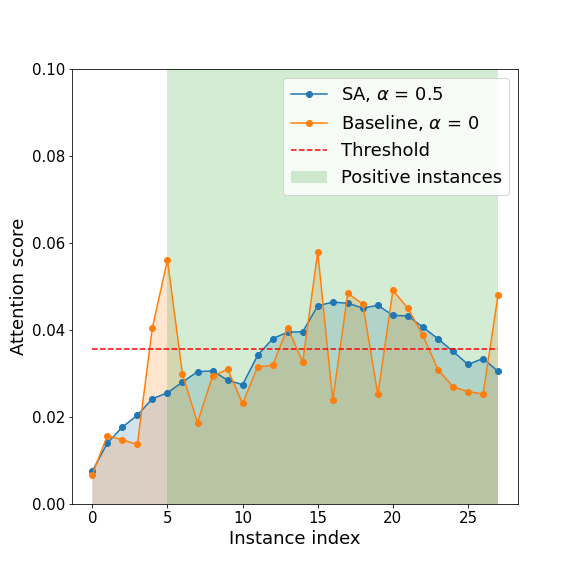}
		}
        \subfloat[Scan 2.]{
                \includegraphics[trim={0.4cm 1cm 2cm 1.3cm},clip,width=0.3\textwidth]{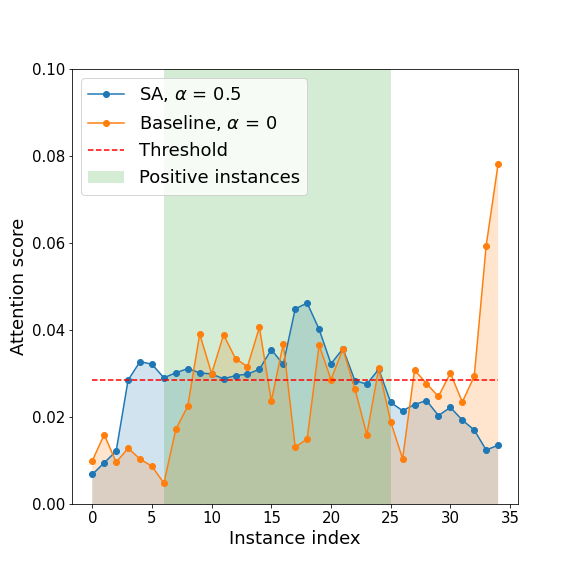}
		}
        \subfloat[Scan 3.]{
                \includegraphics[trim={0.4cm 1cm 2cm 1.3cm},clip,width=0.3\textwidth]{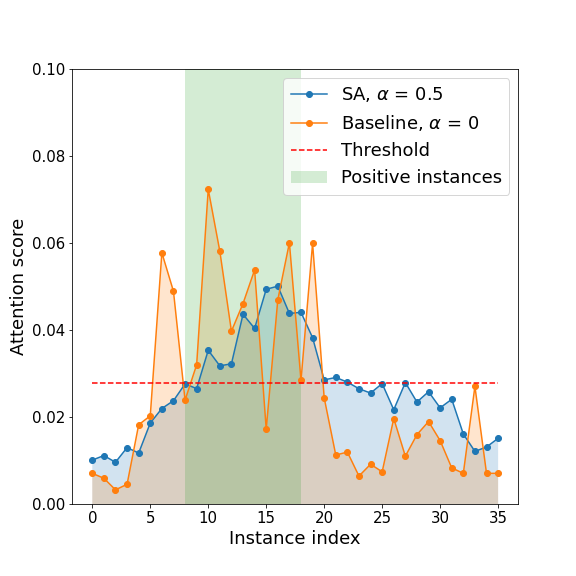}
		}
\vspace*{-3mm}
\caption{Attention weights of SA-DMIL-$S1$ (blue lines, $\alpha = 0.5$) and Att-MIL \cite{ilse2018attention} (orange lines, $\alpha = 0.0$). Slices with values above the threshold ($1/N_b$) are predicted as ICH, while those below are predicted as Normal. The green area highlights those slices whose ground truth label is ICH.}
\label{fig1}
\vspace*{-3mm}
\end{figure}

\vspace*{-1mm} 
So far we have observed enhanced performance through the SA term. In this subsection, we visually illustrate how this novel term imposes smoothness between attention scores of consecutive slices, leading to more accurate predictions. 
Figure \ref{fig1} shows plots of the attention scores assigned by SA-DMIL-$S1$ and Att-MIL to the slices of three different scans (Fig. S1 in the appendix contains an analogous plot for SA-DMIL-$S2$). As expected, introducing the SA loss results in smoother attention weights. Note that the smoothness constraint of SA-DMIL effectively penalizes the appearance of isolated non-smooth attention weights that incorrectly jump over or below the threshold. 

\begin{figure}[ht]
\centering
\resizebox{\textwidth}{!}{%
\begin{tabular}{@{}ccccccc@{}}
\toprule
Scan 1 & \raisebox{-0.5\height}{\includegraphics[trim={0.0cm 0cm 0cm 0cm},clip,width=0.15\textwidth]{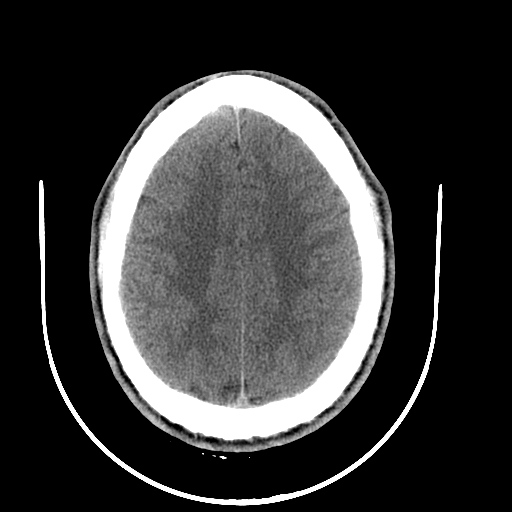}} & \raisebox{-0.5\height}{\includegraphics[trim={0.0cm 0cm 0cm 0cm},clip,width=0.15\textwidth]{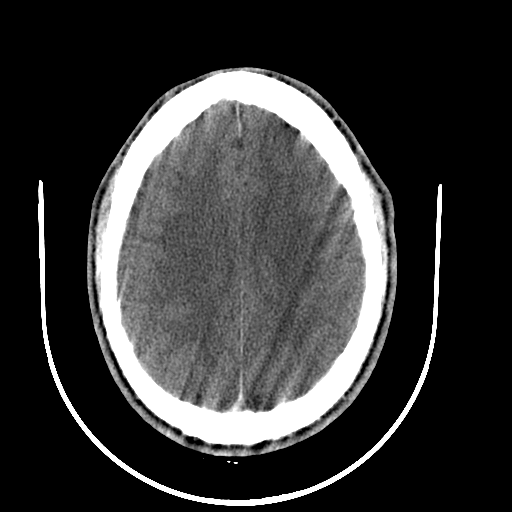}} & \raisebox{-0.5\height}{\includegraphics[trim={0.0cm 0cm 0cm 0cm},clip,width=0.15\textwidth]{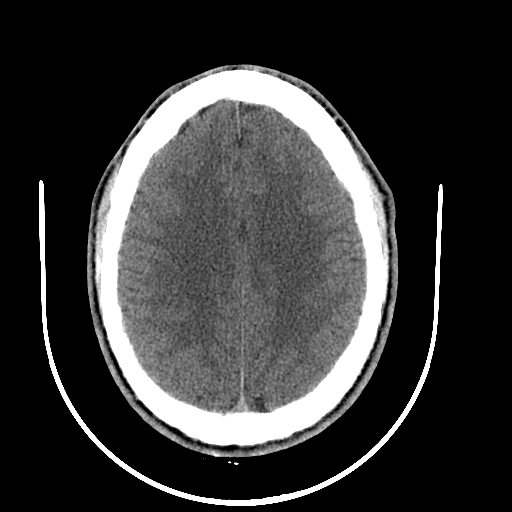}} & \raisebox{-0.5\height}{\includegraphics[trim={0.0cm 0cm 0cm 0cm},clip,width=0.15\textwidth]{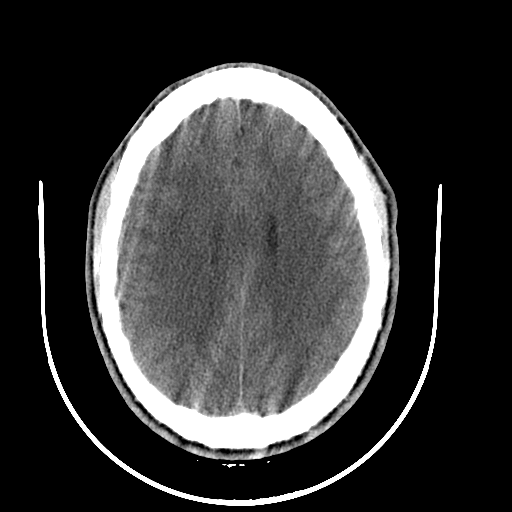}} & \raisebox{-0.5\height}{\includegraphics[trim={0.0cm 0cm 0cm 0cm},clip,width=0.15\textwidth]{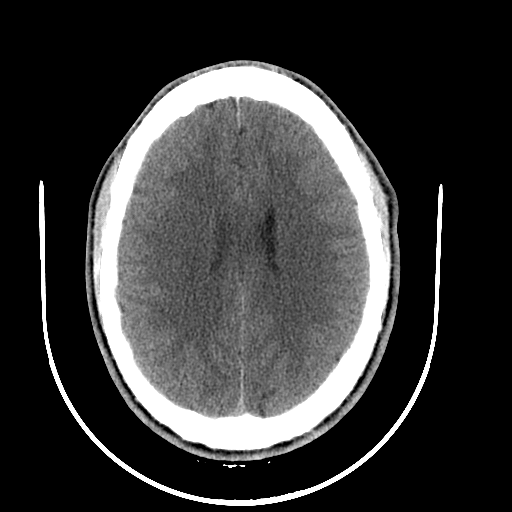}} & \begin{tabular}{c}Scan\\label\end{tabular} \\ \midrule
Ground truth & Normal & Normal & Normal & Normal & Normal & Normal \\
Att-MIL ($\alpha=0.0$) \cite{ilse2018attention} & ICH \tikz\draw[red,fill=red] (0,0) circle (.7ex); & ICH \tikz\draw[red,fill=red] (0,0) circle (.7ex); & Normal \tikz\draw[green,fill=green] (0,0) circle (.7ex); & ICH \tikz\draw[red,fill=red] (0,0) circle (.7ex); & Normal \tikz\draw[green,fill=green] (0,0) circle (.7ex); & ICH \tikz\draw[red,fill=red] (0,0) circle (.7ex); \\
SA-DMIL-$S1$ ($\alpha=0.5$) & Normal \tikz\draw[green,fill=green] (0,0) circle (.7ex); & Normal \tikz\draw[green,fill=green] (0,0) circle (.7ex); & Normal \tikz\draw[green,fill=green] (0,0) circle (.7ex); & Normal \tikz\draw[green,fill=green] (0,0) circle (.7ex); & Normal \tikz\draw[green,fill=green] (0,0) circle (.7ex); & Normal \tikz\draw[green,fill=green] (0,0) circle (.7ex); \\ \midrule \midrule
Scan 2 & \raisebox{-0.5\height}{\includegraphics[trim={0.0cm 0cm 0cm 0cm},clip,width=0.15\textwidth]{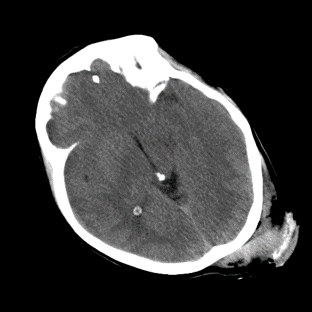}} & \raisebox{-0.5\height}{\includegraphics[trim={0.0cm 0cm 0cm 0cm},clip,width=0.15\textwidth]{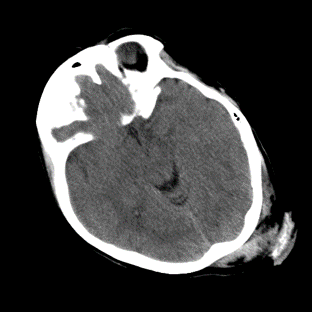}} & \raisebox{-0.5\height}{\includegraphics[trim={0.0cm 0cm 0cm 0cm},clip,width=0.15\textwidth]{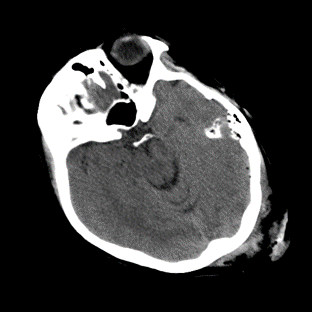}} & \raisebox{-0.5\height}{\includegraphics[trim={0.0cm 0cm 0cm 0cm},clip,width=0.15\textwidth]{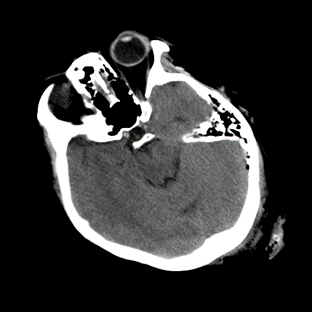}} & \raisebox{-0.5\height}{\includegraphics[trim={0.0cm 0cm 0cm 0cm},clip,width=0.15\textwidth]{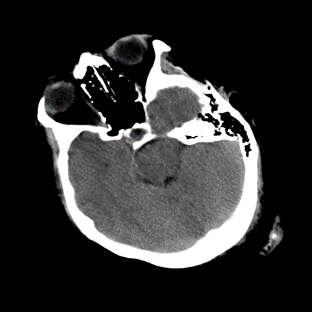}} & \begin{tabular}{c}Scan\\label\end{tabular} \\ \midrule
Ground truth & ICH & ICH & Normal & Normal & Normal & ICH  \\
Att-MIL ($\alpha=0.0$) \cite{ilse2018attention} & Normal \tikz\draw[red,fill=red] (0,0) circle (.7ex); & ICH \tikz\draw[green,fill=green] (0,0) circle (.7ex); & ICH \tikz\draw[red,fill=red] (0,0) circle (.7ex); & ICH \tikz\draw[red,fill=red] (0,0) circle (.7ex); & Normal \tikz\draw[green,fill=green] (0,0) circle (.7ex); & ICH \tikz\draw[green,fill=green] (0,0) circle (.7ex);\\
SA-DMIL-$S1$ ($\alpha=0.5$) & ICH \tikz\draw[green,fill=green] (0,0) circle (.7ex); & ICH \tikz\draw[green,fill=green] (0,0) circle (.7ex); & Normal \tikz\draw[green,fill=green] (0,0) circle (.7ex); & Normal \tikz\draw[green,fill=green] (0,0) circle (.7ex); & Normal \tikz\draw[green,fill=green] (0,0) circle (.7ex); & ICH \tikz\draw[green,fill=green] (0,0) circle (.7ex); \\ \bottomrule
\end{tabular}%
}
\vspace{-2mm}
\caption{Predictions of Att-MIL \cite{ilse2018attention} and SA-DMIL-$S1$ at CT slice level in two different scans. SA improves predictions at both scan and slice level. Red color: incorrect prediction, green color: correct prediction.}
\label{table:show_ex}
\vspace{-3mm}
\end{figure}

We also include visual examples of consecutive CT slices in Fig. \ref{table:show_ex}. In Scan 1, the baseline Att-MIL produces a wrong prediction at scan level. When using SA, the prediction is correct since dependencies between adjacent slices have been learned. In Scan 2, both models produce correct predictions at scan level, but SA-DMIL is more accurate at slice level. This occurs thanks to the SA loss, that turns the attention scores into smoother values and, therefore, avoids random \emph{jumps} up and down the decision threshold.

\vspace{-3mm}
\section{Conclusion}
\vspace{-1mm}

In this study we have proposed SA-DMIL, a new model that obtains significant improvements in ICH classification compared to state-of-the-art MIL methods. 
This is done by adding a smoothing regularizing term to the loss function. 
This term imposes a smoothness constraint on the latent function that encodes the attention weights, which forces our model to learn dependencies between instances rather than training each instance independently in a bag. 
This flexible approach does not introduce any additional complexity, so similar ideas can be applied to other methods to model dependencies between neighboring instances. 


\vspace{-2mm}
\section*{Data use declaration}
\vspace{-2mm}
The dataset used in this study is from the 2019 RSNA Intracranial Hemorrhage Detection Challenge and is publicly available \href{https://www.kaggle.com/competitions/rsna-intracranial-hemorrhage-detection/data}{in this link}.

\section*{Acknowledgements}

This work was supported by project PID2019-105142RB-C22 funded by Ministerio de Ciencia e Innovación and by project B-TIC-324-UGR20 funded by FEDER/Junta de Andalucía and Universidad de Granada. The work by Francisco M. Castro-Macías was supported by Ministerio de Universidades under FPU contract FPU21/01874.

\bibliographystyle{splncs04}
\bibliography{bib}

\newpage
\setcounter{page}{1}
\appendix
\section*{Supplementary Materials}

\renewcommand{\thetable}{S\arabic{table}}
\setcounter{table}{0}

\begin{table}[ht]
\caption{Performance of SA-DMIL and other MIL methods at scan level on the RSNA dataset. The average of 5 independent runs is reported.}
\centering
\resizebox{\textwidth}{!}{%
\begin{tabular}{@{}cc|ccccc@{}}
\toprule
\multicolumn{2}{c|}{} & \multicolumn{5}{c}{Scan level} \\ 
\multicolumn{2}{c|}{Model} & Acc & Pre & Rec & F1 & AUC \\ \midrule
\multirow{9}{*}{SA-DMIL-$S1$} & $\alpha=0.9$ & 0.753 $\pm$ 0.014 & 0.803 $ \pm$0.006 & 0.681 $\pm$ 0.017 & 0.735 $\pm$ 0.011 & 0.839 $\pm$ 0.007 \\
 & $\alpha=0.8$ & 0.754 $\pm$ 0.010 & 0.801 $\pm$ 0.004 & 0.713 $\pm$ 0.014 & 0.740 $\pm$ 0.009 & 0.851 $\pm$ 0.006 \\
 & $\alpha=0.7$ & 0.806 $\pm$ 0.007 & 0.763 $\pm$ 0.008 & 0.784 $\pm$ 0.010 & 0.775 $\pm$ 0.008 & 0.860 $\pm$ 0.010 \\
 & $\alpha=0.6$ & 0.801 $\pm$ 0.009 & 0.791 $\pm$ 0.004 & 0.806 $\pm$ 0.009 & 0.799 $\pm$ {0.006} & 0.869 $\pm$ 0.007 \\
 & $\alpha=0.5$ & \textbf{0.813} $\pm$ {0.010} & 0.805 $\pm$ {0.003} & 0.806 $\pm$ 0.011 & \textbf{0.806} $\pm$ {0.006} & 0.879 $\pm$ {0.003} \\
 & $\alpha=0.4$ & 0.773 $\pm$ 0.015 & 0.806 $\pm$ 0.007 & 0.694 $\pm$ 0.018 & 0.746 $\pm$ 0.012 & 0.863 $\pm$ 0.004 \\
 & $\alpha=0.3$ & 0.767 $\pm$ 0.012 & 0.734 $\pm$ 0.011 & 0.806 $\pm$ 0.007 & 0.768 $\pm$ 0.008 & 0.859 $\pm$ 0.010 \\
 & $\alpha=0.2$ & 0.747 $\pm$ 0.021 & 0.681 $\pm$ 0.010 & \textbf{0.889} $\pm$ {0.004} & 0.771 $\pm$ 0.006 & 0.857 $\pm$ 0.011 \\
 & $\alpha=0.1$ & 0.747 $\pm$ 0.017 & 0.783 $\pm$ 0.010 & 0.652 $\pm$ 0.018 & 0.712 $\pm$ 0.014 & 0.841 $\pm$ 0.014 \\ \midrule
\multirow{9}{*}{SA-DMIL-$S2$} & $\alpha=0.9$ & 0.753 $\pm$ 0.012 & 0.817 $\pm$ 0.010 & 0.613 $\pm$ 0.019 & 0.714 $\pm$ 0.013 & 0.816 $\pm$ 0.010 \\
 & $\alpha=0.8$ & 0.733 $\pm$ 0.021 & 0.813 $\pm$ 0.008 & 0.656 $\pm$ 0.017 & 0.747 $\pm$ 0.009 & 0.825 $\pm$ 0.008 \\
 & $\alpha=0.7$ & 0.767 $\pm$ 0.014 & 0.776 $\pm$ 0.012 & 0.722 $\pm$ 0.013 & 0.748 $\pm$ 0.010 & 0.843 $\pm$ 0.008 \\
 & $\alpha=0.6$ & 0.760 $\pm$ 0.009 & 0.846 $\pm$ 0.007 & 0.711 $\pm$ 0.014 & 0.778 $\pm$ 0.008 & 0.852 $\pm$ 0.007 \\
 & $\alpha=0.5$ & 0.800 $\pm$ 0.008 & 0.828 $\pm$ 0.010 & 0.736 $\pm$ 0.017 & 0.780 $\pm$ 0.011 & 0.867 $\pm$ 0.008 \\
 & $\alpha=0.4$ & 0.773 $\pm$ 0.010 & \textbf{0.896} $\pm$ 0.005 & 0.697 $\pm$ 0.013 & 0.737 $\pm$ 0.007 & \textbf{0.880} $\pm$ 0.004 \\
 & $\alpha=0.3$ & 0.763 $\pm$ 0.012 & 0.797 $\pm$ 0.007 & 0.686 $\pm$ 0.016 & 0.721 $\pm$ 0.010 & 0.853 $\pm$ 0.011 \\
 & $\alpha=0.2$ & 0.753 $\pm$ 0.010 & 0.818 $\pm$ 0.006 & 0.625 $\pm$ 0.021 & 0.709 $\pm$ 0.017 & 0.838 $\pm$ 0.012 \\
 & $\alpha=0.1$ & 0.747 $\pm$ 0.019 & 0.736 $\pm$ 0.011 & 0.740 $\pm$ 0.012 & 0.736 $\pm$ 0.013 & 0.833 $\pm$ 0.016 \\ \midrule
\multicolumn{2}{c|}{Att-MIL ($\alpha = 0.0$) \cite{ilse2018attention}} & 0.740 $\pm$ 0.015 & 0.674 $\pm$ 0.024 & 0.832 $\pm$ 0.011 & 0.719 $\pm$ 0.014 & 0.829 $\pm$ 0.009 \\
\multicolumn{2}{c|}{MIL + Max agg. \cite{wang2019comparison}} & 0.617 $\pm$ 0.031 & 0.856 $\pm$ 0.030 & 0.447 $\pm$ 0.109 & 0.575 $\pm$ 0.068 & 0.743 $\pm$ 0.015 \\
\multicolumn{2}{c|}{MIL + Mean agg.\cite{wang2019comparison}} & 0.677 $\pm$ 0.028 & 0.670 $\pm$ 0.032 & 0.734 $\pm$ 0.041 & 0.693 $\pm$ 0.040 & 0.801 $\pm$ 0.016 \\
\multicolumn{2}{c|}{Att-CNN + VGPMIL \cite{wu2021combining}} & 0.765 $\pm$ 0.017 & 0.724 $\pm$ 0.012 & 0.851 $\pm$ 0.008 & 0.773 $\pm$ 0.010 & 0.868 $\pm$ 0.007 \\ \bottomrule
\end{tabular}%
}
\label{table-S1}
\end{table}

\begin{table}[ht]
\caption{Performance of SA-DMIL and other MIL methods at slice level on the RSNA dataset. The average of 5 independent runs is reported.}
\centering
\resizebox{\textwidth}{!}{%
\begin{tabular}{@{}cc|cccc@{}}
\toprule
\multicolumn{2}{c|}{} & \multicolumn{4}{c}{Slice level} \\ 
\multicolumn{2}{c|}{Model} & Acc & Pre & Rec & F1 \\ \midrule
\multirow{9}{*}{SA-DMIL-$S1$} & $\alpha=0.9$ & 0.789 $\pm$ 0.025 & 0.670 $\pm$ 0.031 & 0.541 $\pm$ 0.051 & 0.598 $\pm$ 0.048 \\
 & $\alpha=0.8$ & 0.794 $\pm$ 0.017 & 0.683 $\pm$ 0.027 & 0.548 $\pm$ 0.047 & 0.622 $\pm$ 0.040 \\
 & $\alpha=0.7$ & 0.828 $\pm$ 0.011 & 0.679 $\pm$ 0.028 & 0.576 $\pm$ 0.038 & 0.639 $\pm$ 0.037 \\
 & $\alpha=0.6$ & 0.821 $\pm$ 0.014 & 0.687 $\pm$ 0.025 & 0.579 $\pm$ 0.037 & 0.643 $\pm$ 0.029 \\
 & $\alpha=0.5$ & \textbf{0.834} $\pm$ 0.010 & 0.732 $\pm$ 0.021 & \textbf{0.608} $\pm$ 0.027 & \textbf{0.686} $\pm$ 0.018 \\
 & $\alpha=0.4$ & 0.788 $\pm$ 0.017 & 0.718 $\pm$ 0.023 & 0.563 $\pm$ 0.031 & 0.647 $\pm$ 0.024 \\
 & $\alpha=0.3$ & 0.775 $\pm$ 0.018 & 0.702 $\pm$ 0.028 & 0.551 $\pm$ 0.036 & 0.624 $\pm$ 0.032 \\
 & $\alpha=0.2$ & 0.768 $\pm$ 0.023 & 0.661 $\pm$ 0.034 & 0.551 $\pm$ 0.032 & 0.611 $\pm$ 0.035 \\
 & $\alpha=0.1$ & 0.766 $\pm$ 0.022 & 0.649 $\pm$ 0.032 & 0.540 $\pm$ 0.041 & 0.584 $\pm$ 0.039 \\ \midrule
\multirow{9}{*}{SA-DMIL-$S2$} & $\alpha=0.9$ & 0.768 $\pm$ 0.030 & 0.733 $\pm$ 0.024 & 0.551 $\pm$ 0.048 & 0.598 $\pm$ 0.046 \\
 & $\alpha=0.8$ & 0.766 $\pm$ 0.022 & 0.736 $\pm$ 0.022 & 0.573 $\pm$ 0.046 & 0.625 $\pm$ 0.042 \\
 & $\alpha=0.7$ & 0.807 $\pm$ 0.018 & 0.734 $\pm$ 0.028 & 0.591 $\pm$ 0.049 & 0.638 $\pm$ 0.039 \\
 & $\alpha=0.6$ & 0.801 $\pm$ 0.020 & 0.746 $\pm$ 0.026 & 0.594 $\pm$ 0.037 & 0.655 $\pm$ 0.029 \\
 & $\alpha=0.5$ & 0.823 $\pm$ 0.012 & 0.748 $\pm$ 0.022 & 0.596 $\pm$ 0.029 & 0.659 $\pm$ 0.024 \\
 & $\alpha=0.4$ & 0.812 $\pm$ 0.017 & \textbf{0.751} $\pm$ 0.020 & 0.583 $\pm$ 0.027 & 0.637 $\pm$ 0.019 \\
 & $\alpha=0.3$ & 0.790 $\pm$ 0.021 & 0.738 $\pm$ 0.028 & 0.561 $\pm$ 0.034 & 0.622 $\pm$ 0.031 \\
 & $\alpha=0.2$ & 0.784 $\pm$ 0.024 & 0.742 $\pm$ 0.024 & 0.551 $\pm$ 0.037 & 0.617 $\pm$ 0.028 \\
 & $\alpha=0.1$ & 0.767 $\pm$ 0.031 & 0.683 $\pm$ 0.029 & 0.547 $\pm$ 0.040 & 0.593 $\pm$ 0.037 \\ \midrule
\multicolumn{2}{c|}{Att-MIL ($\alpha = 0.0$) \cite{ilse2018attention}} & 0.751 $\pm$ 0.024 & 0.623 $\pm$ 0.037 & 0.543 $\pm$ 0.047 & 0.579 $\pm$ 0.041 \\
\multicolumn{2}{c|}{MIL + Max agg. \cite{wang2019comparison}} & 0.732 $\pm$ 0.041 & 0.441 $\pm$ 0.108 & 0.373 $\pm$ 0.152 & 0.406 $\pm$ 0.128 \\
\multicolumn{2}{c|}{MIL + Mean agg. \cite{wang2019comparison}} & 0.741 $\pm$ 0.038 & 0.502 $\pm$ 0.110 & 0.386 $\pm$ 0.117 & 0.447 $\pm$ 0.107 \\
\multicolumn{2}{c|}{Att-CNN + VGPMIL \cite{wu2021combining}} & 0.807 $\pm$ 0.022 & 0.714 $\pm$ 0.028 & 0.538 $\pm$ 0.039 & 0.597 $\pm$ 0.036 \\ \bottomrule
\end{tabular}%
}
\label{table-S2}
\end{table}

\renewcommand{\thefigure}{S\arabic{figure}}
\setcounter{figure}{0}

\begin{figure}[ht]
\centering
    \subfloat[Bag 1.]{
                \includegraphics[trim={0.4cm 1cm 2cm 1.3cm},clip,width=0.3\textwidth]{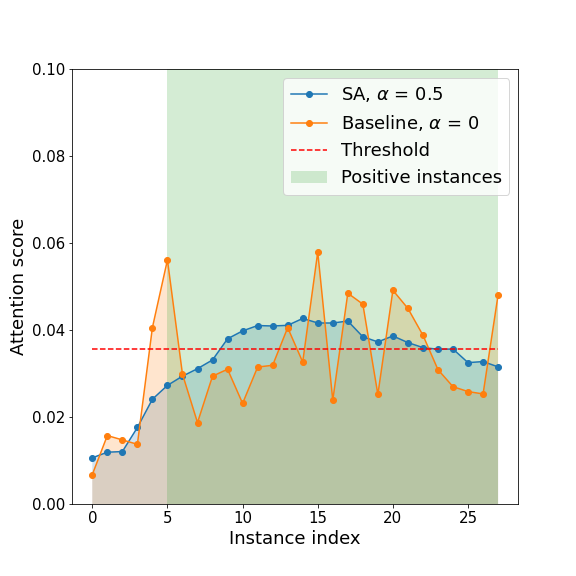}
        }
        \subfloat[Bag 2.]{
                \includegraphics[trim={0.4cm 1cm 2cm 1.3cm},clip,width=0.3\textwidth]{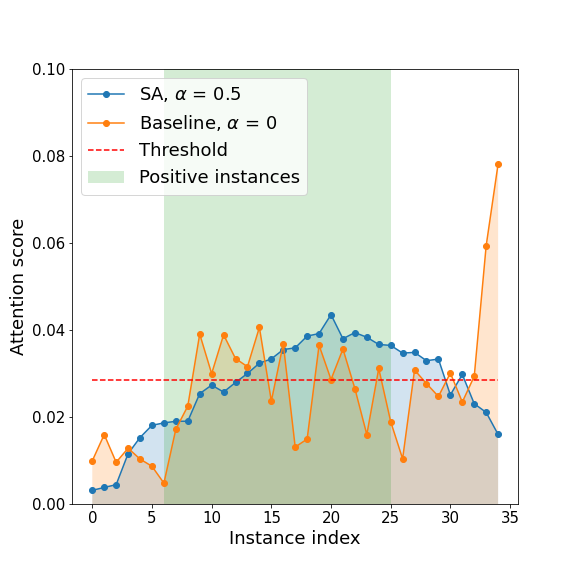}
        }
        \subfloat[Bag 3.]{
                \includegraphics[trim={0.4cm 1cm 2cm 1.3cm},clip,width=0.3\textwidth]{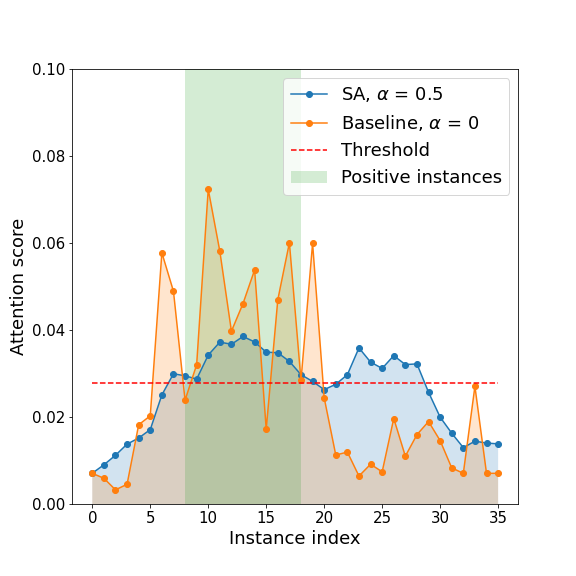}
        }
\caption{Attention weights of SA-DMIL-$S2$ (blue lines, $\alpha = 0.5$) and Att-MIL \cite{ilse2018attention} (orange lines, $\alpha = 0.0$). Slices with values above the threshold ($1/N_b$) are predicted as ICH, while those below are predicted as Normal. The green area highlights those slices whose ground truth label is ICH.} 
\label{fig2}
\end{figure}

\end{document}